\documentclass[graybox]{svmult}

\usepackage{mathptmx}       
\usepackage{helvet}         
\usepackage{courier}        
\usepackage{type1cm}        
\usepackage{makeidx}         
\usepackage{graphicx}        
\usepackage{multicol}        
\usepackage[bottom]{footmisc}

\makeindex             

\begin{document}

\title*{The Black Hole Uncertainty Principle Correspondence}
\author{B. J. Carr}
\institute{B. J. Carr \at Department of Physics and Astronomy, Queen Mary University of London, 
Mile End Rd, 
London E1 4NS, UK,
\email{B.J.Carr@qmul.ac.uk}}
%
%
\maketitle

\abstract{The Black Hole Uncertainty Principle correspondence proposes a connection between the Uncertainty Principle on microscopic scales and black holes on macroscopic scales. This is manifested in a unified expression for the Compton wavelength and Schwarzschild radius.
It is a natural consequence of the Generalized Uncertainty Principle, which suggests corrections to the Uncertainty Principle as the energy increases towards the Planck value. It also entails corrections to the event horizon size as the black hole mass falls to the Planck value, leading to the concept of a Generalized Event Horizon.  One implication of this is that there could be sub-Planckian black holes with a size of order their Compton wavelength. 
Loop quantum gravity suggests the existence of black holes with precisely this feature.
The correspondence leads to a heuristic derivation of the black hole temperature and suggests how the Hawking formula is modified in the sub-Planckian regime.
}

\section{Introduction}
\label{sec:1}

\begin{figure}[b]
\sidecaption
\includegraphics[scale=.35]{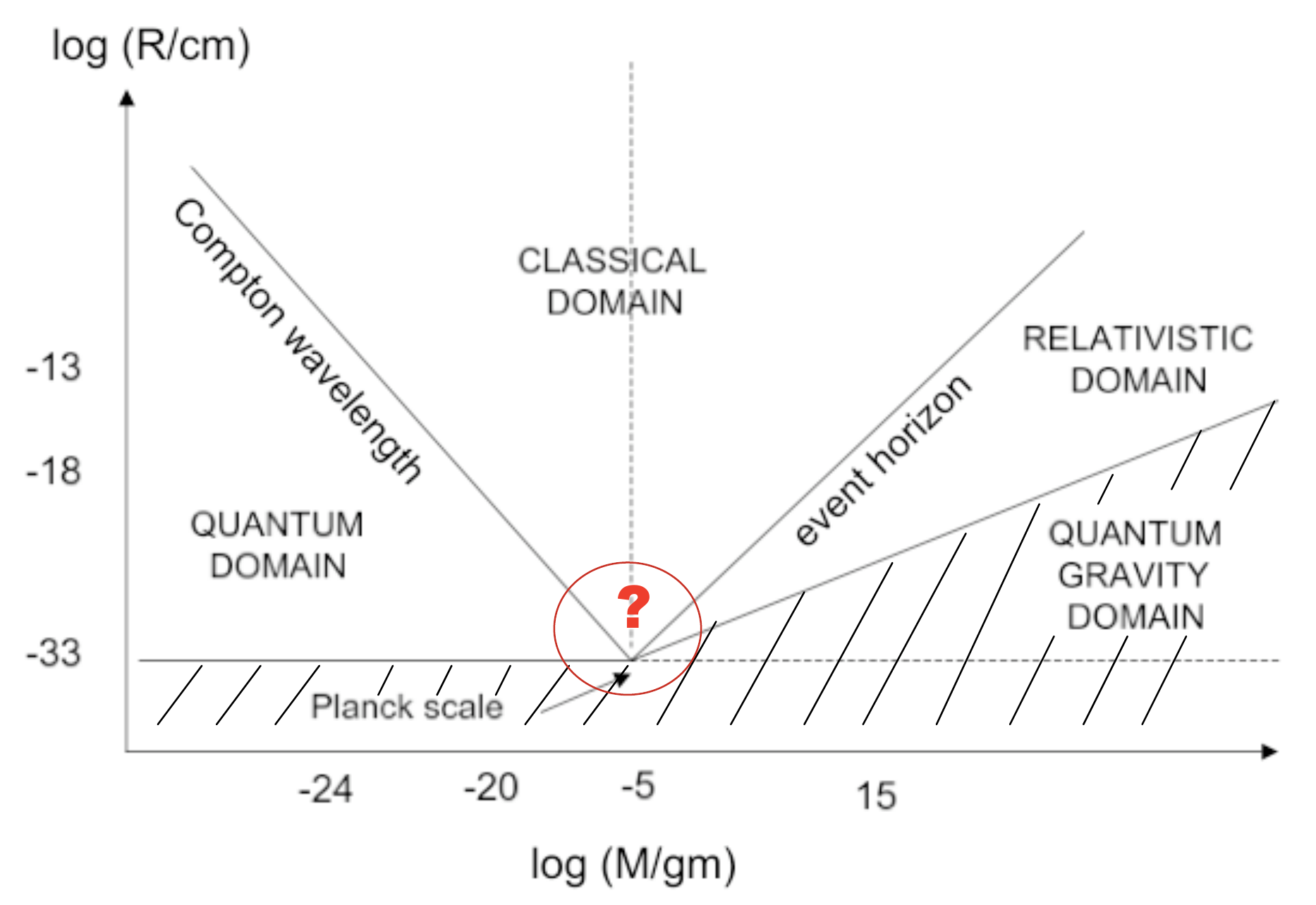}
\caption{The division of the ($M,R$) diagram into the classical, quantum, relativistic and quantum gravity domains.} 
\label{MR}       
\end{figure}

A key feature of the microscopic domain is the
Heisenberg Uncertainty Principle (HUP) which implies that the uncertainty in the position and momentum of a particle must satisfy
$ \Delta x > \hbar/ (2 \Delta p)$.
It is well known that one can heuristically understand this result as reflecting the momentum transferred to the particle by the probing photon.  Since the momentum of a particle of mass $M$ is bounded by 
$Mc$, an immediate implication is that  one cannot localize a particle of mass $M$ on a scale less $\hbar/(2Mc)$. 
An important role is therefore played by the reduced Compton wavelength,
$R_C = \hbar/(Mc)$, 
which can be obtained from the HUP with the substitution $\Delta x \rightarrow R$ and $\Delta p \rightarrow c M$ but without the factor of $2$. In the $(M,R)$ diagram of Fig.~\ref{MR}, the region corresponding to $R<R_C$ 
might  be regarded as the ``quantum domain'' in the sense that the classical description
 breaks down there. 
A key feature of the macroscopic domain is the existence of black holes. 
General relativity implies that a spherically symmetric object of mass $M$ forms an event horizon if it falls within its Schwarzschild radius,
$R_S = 2GM/c^2$.
The region $R<R_S$ might be regarded as the ``relativistic domain" in the sense that there is no stable classical configuration in this part of Fig.~\ref{MR}. 

The Compton and Schwarzschild lines
  intersect at around the Planck scales,
$ R_P = \sqrt{ \hbar G/c^3} \sim 10^{-33} \mathrm {cm}, 
 M_P = \sqrt{ \hbar c/G} \sim 10^{-5} \mathrm g $,
and they divide the $(M,R)$ diagram in Fig.~\ref{MR} into three regimes (quantum, relativistic, classical). There are several other interesting lines in this diagram. The vertical line 
$M=M_P$ is often assumed to mark the division between  elementary particles ($M <M_P$) and black holes ($M > M_P$), because one usually requires a black hole to be larger than its own Compton wavelength.
The horizontal line $R=R_P$ 
is significant because quantum fluctuations in the metric should  become important below this.
Quantum gravity effects should also be important whenever the density exceeds the Planck value,
$\rho_P = c^5/(G^2  \hbar) \sim 10^{94} \mathrm {g \, cm^{-3}}$,
corresponding to the sorts of  curvature singularities associated with the big bang or the centres of  black holes. This implies $R <( M/M_P)^{1/3}R_P $,
which is well above the $R = R_P$ line in Fig.~\ref{MR} for $M \gg M_P$, so one might regard the shaded region as specifying the ``quantum gravity" domain.

\begin{figure}[b]
\sidecaption
\includegraphics[scale=.3]{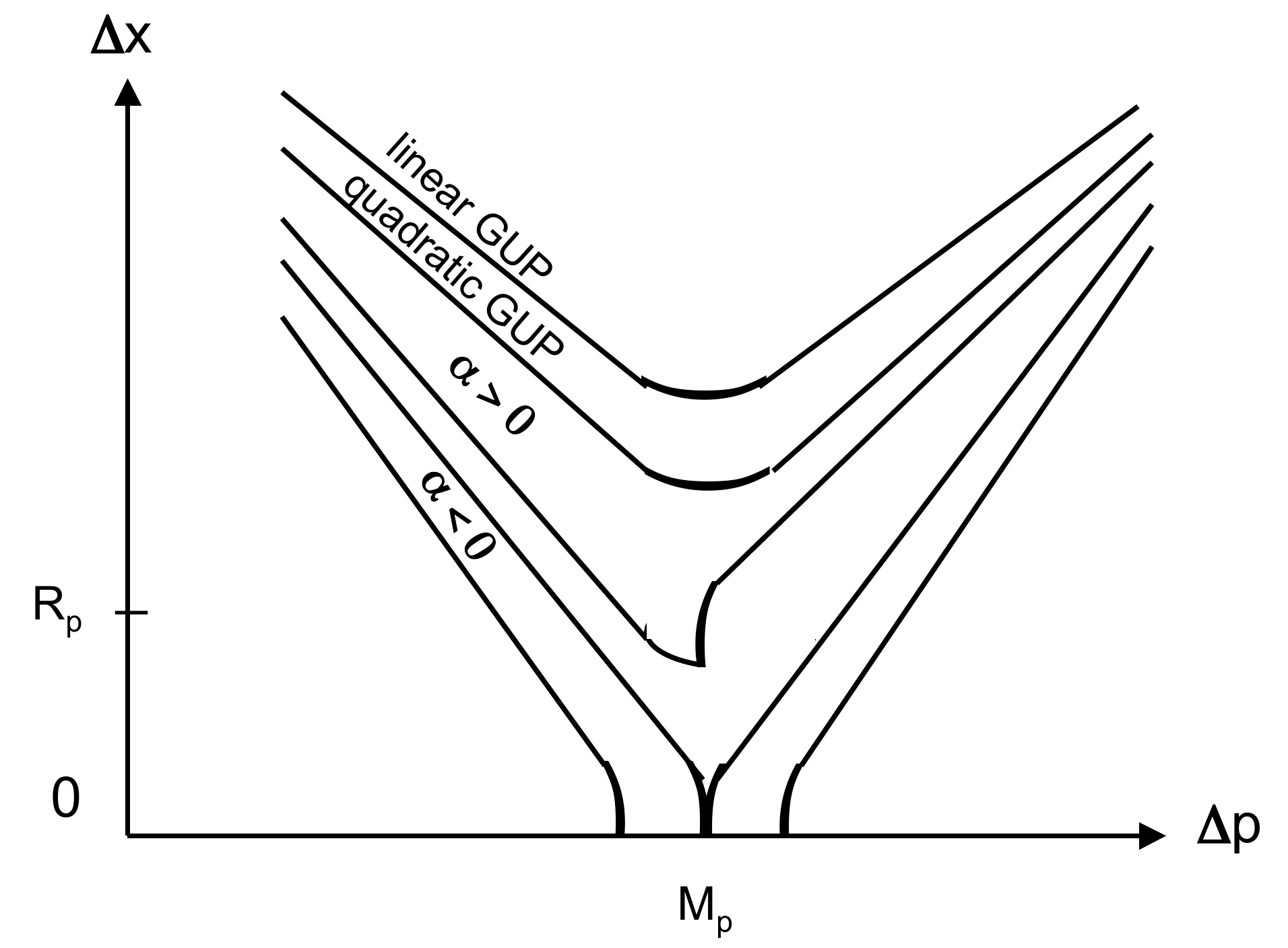}
\caption{$\Delta x$ versus $\Delta p$ for the GUP in its linear and quadratic forms and for more exotic possibilities.}
\label{modesto1}       
\end{figure}

Although the Compton and Schwarzschild boundaries 
correspond to straight lines in the logarithmic plot of Fig.~\ref{MR}, this form presumably breaks down near the Planck point. As one approaches the Planck point from the left, Adler~\cite{Adler} and many others have argued 
that the HUP should be replaced by a Generalized Uncertainty Principle (GUP) of the form
 \begin{equation}
 \Delta x > \hbar/ \Delta p + \alpha R_P^2 (\Delta p/ \hbar) \, .
\label{GUP1} 
\end{equation}
Here $\alpha$ is a dimensionless constant (usually assumed positive) which depends on the particular model and the factor of $2$ in the first term has been dropped. A heuristic argument for the second term in Eq.~(\ref{GUP1}) is that it represents the gravitational effect  of the probing photon rather than its momentum effect. This form of the GUP is indicated by the top curve in
Fig.~\ref{modesto1}.
Variants of Eq.~(\ref{GUP1}) can be found in other approaches to quantum gravity, such as
non-commutative quantum mechanics or general minimum length considerations \cite{maggiore}. 
The GUP can also be derived in loop quantum gravity 
because of polymer corrections in the structure of spacetime \cite{ashtekar} and it is implicit in some approaches to the problem of quantum decoherence~\cite{kay}. An expression resembling Eq.~(\ref{GUP1}) also arises in string theory~\cite{veneziano}.

The second term on the right of Eq.~(\ref{GUP1}) is much smaller than the first term for $\Delta p \ll M_Pc$. Since it can be written as $\alpha G (\Delta p)/c^3$, it roughly  corresponds to the Schwarzschild radius for an object of mass $\Delta p/c$.  
 Indeed, if we rewrite Eq.~(\ref{GUP1}) using the same substitution $\Delta x \rightarrow R$ and $\Delta p \rightarrow c M$ as before, 
it becomes
 \begin{equation}
R > R_C' =  \hbar/(Mc) + \alpha GM/c^2 = \frac{\hbar}{Mc} \left[ 1  + \alpha (M/M_P)^2 \right]  \, .
\label{GUP2}
 \end{equation}
The lower limit on $R$ might be regarded as a generalized Compton wavelength,
the last term representing a small correction as one approaches the Planck point from the left.
However, one can also apply Eq.~(\ref{GUP2}) for $M \gg M_P$ and it is interesting that in this regime it asymptotes to the Schwarzschild 
form, apart from a numerical factor \cite{cmp}. This suggests that  there is a different kind of positional uncertainty for an object larger than the Planck mass, related to the existence of black holes. 
This is not unreasonable since the Compton wavelength is below the Planck scale (and hence meaningless) here and also an outside observer cannot localize an object on a scale smaller than its Schwarzschild radius. 

The GUP also has important implications for the black hole horizon size, as can be seen by  examining what happens as one approaches the intersect point from the right. In this limit, 
it is natural to write Eq.~(\ref{GUP2}) as
 \begin{equation}
R > R_S' = \frac{\alpha GM}{c^2} \left[ 1  + \frac{1}{\alpha} (M_P/ M)^2 \right] 
\label{GEH2}
\end{equation}
and this represents a small perturbation to the Schwarzschild radius for $M \gg M_P$ 
if one assumes $\alpha =2$. 
However, there is no reason for anticipating $\alpha=2$ in the heuristic derivation of the GUP. Nor is it clear why a more  precise calculation (within the context of a specific theory of quantum gravity) would yield this value. 

This motivates an alternative approach in which the free constant in Eq.~(\ref{GUP2}) is associated with the first term rather than the second. After all, the factor of 2 in the expression for the Schwarzschild radius is precise, whereas  the coefficient associated with the Compton term is somewhat arbitrary. 
Thus one might replace Eqs.~(\ref{GUP2}) and (\ref{GEH2}) with the expressions
 \begin{equation}
R_C' = 
\frac{\beta \hbar}{Mc} \left[ 1  + \frac{2}{\beta} (M/M_P)^2 \right] , \quad
R_S' = \frac{2 GM}{c^2} \left[ 1  + \frac{\beta}{2} (M_P/ M)^2 \right]  \,.
\label{GUP4A}
\end{equation}
for some constant $\beta$, with
the second expression being regarded as a Generalized Event Horizon (GEH). The mathematical equivalence of $R_C'$ and $R_S'$ underlies what we have termed the BHUP correspondence.

An important caveat is that
Eq.~(\ref{GUP1}) assumes the two uncertainties add linearly.
 On the other hand, 
since they are independent, it might be more natural to assume that they add quadratically:
 \begin{equation}
 \Delta x > \sqrt {(\hbar/ \Delta p)^2 + (\alpha R_P^2 \Delta p/ \hbar)^2} \, .
 \label{quad2}
 \end{equation}
This corresponds to the lower GUP curve in Fig.~\ref{modesto1}. While the heuristic arguments indicate the form of the two uncertainty terms, they do not specify how one combines them. We refer to Eqs.~(\ref{GUP1}) and (\ref{quad2}) as the {\it linear} and {\it quadratic} forms of the GEP. 
Adopting the $\beta$ formalism, as before, then gives a unified expression for  generalized Compton wavelength and  event horizon size 
\begin{equation}
R_C' = R_S' = \sqrt{ (\beta \hbar /Mc)^2  + (2GM/c^2)^2 } \, ,
\label{quadEH}
\end{equation}
leading to the approximations
 \begin{equation}
R _C' \approx \frac{\beta \hbar}{Mc}  \left[ 1  + \frac{2}{\beta^2} (M/M_P)^4 \right] , \quad
R _S' \approx \frac{2GM}{c^2} \left[ 1  + \frac{\beta^2}{8} (M_P/ M)^4 \right] 
\label{quadC}
\end{equation}
for $M \ll M_P$ and $M \gg M_P$, respectively. These might be compared to the {\it exact} expressions in the linear case, given by Eq.~(\ref{GUP4A}).
As shown below, the horizon size of the black hole solution in loop quantum gravity has precisely the form (\ref{quadEH}).

More generally, the BHUP correspondence might allow any unified expression for $R_C'(M) \equiv R_S'(M)$ which has the asymptotic behaviour
$\beta \hbar/(Mc)$ for $M \ll M_P$ and 
$2GM/c^2$ for $M \gg M_P$.
The continuity between the Compton and Schwarzschild lines in  Fig.~\ref{modesto1} might then suggest some link between elementary particles and black holes \cite{cmp}. It may also relate to the 
``firewall'' proposal \cite{fire} (i.e. some new quantum effect at the black hole horizon).  
However, the distinction between between black holes and elementary particles could be maintained in models \cite{nicolini} with some form of discontinuity at the minimum (middle curve) or in models
with $\alpha < 0$ (bottom curves). One might even consider models with a cusp, such that $G \rightarrow 0$  (no gravity) and  $\hbar \rightarrow 0$ (no quantum discreteness) at the minimum.
This relates to papers presented  at this meeting on ``asymptotic safety'' \cite{bonnano} and ``world crystals'' \cite{scardigli}. 

\section{Loop Black Holes}

Loop quantum gravity
(LQG) is based on a canonical quantization of the Einstein equations written in terms of the Ashtekar variables~\cite{LQGgeneral}.
One important consequence is 
that the area is quantized, with the smallest possible 
value given by
\begin{equation}
a_o \equiv A_{\rm min}/8\pi= \sqrt{3}\, \gamma\zeta R_P^2/2 \, ,
\end{equation}
where $\gamma$ is the Immirzi parameter and 
$\zeta$ is another constant, both being 
of order $1$.
The other relevant constant is the dimensionless polymeric
parameter  $\delta$, which (together with $a_0$) determines the deviation from classical theory.

One version of LQG
gives a black hole solution, known as the loop black hole (LBH)
 \cite{poly}, which exhibits self-duality and
replaces the singularity with
another asymptotically flat region. 
The metric in this solution depends only on the dimensionless parameter $\epsilon \equiv \delta \gamma$, which must be small, and
can be expressed as \cite{poly}
\begin{eqnarray}
ds^2 = - G(r) c^2 dt^2 + dr^2/F(r) + H(r) (d \theta^2 + \sin^2 \theta d \phi^2) \, ,
\label{g}
\end{eqnarray}
\begin{equation}
H = r^2 + \frac{a_o^2}{r^2}, \,
G = \frac{(r-r_+)(r-r_-)(r+ r_{*})^2}{r^4 +a_o^2}, \,
F = \frac{(r-r_+)(r-r_-) r^4}{(r+ r_{*})^2 (r^4 +a_o^2)}. \nonumber 
\end{equation}
Here $r_+ = 2Gm/c^2$ and $r_-= 2G m P^2/c^2$ are the outer and inner horizons, respectively, and $r_* \equiv \sqrt{r_+ r_-} = 2GmP/c^2$, where $m$ is the black hole mass and 
\begin{equation}
P \equiv  \frac {\sqrt{1+\epsilon^2} -1}{\sqrt{1+\epsilon^2} +1}  
\end{equation}
is the polymeric function. 
For $\epsilon \ll 1$, we have $P \approx \epsilon^2 /4 \ll 1$ , so $r_- \ll r_* \ll r_+$. 
 
In the limit $r \to \infty$, $H(r) \approx r^2$, so $r$ is the usual radial coordinate and $F(r) \approx G(r) \approx 1-2 G M/(c^2 r)$ where 
$M = m (1+P)^2$
is the ADM mass.
However, the exact expression for $H(r)$ shows that the physical radial coordinate
$R = \sqrt{H}$
decreases from $\infty$ to a minimum $\sqrt{2a_0}$ at $r=\sqrt{a_0}$ and then increases again to $\infty$ as $r$ decreases from $\infty$ to $0$. 
In particular, the value of $R$ associated with the event horizon is
\begin{eqnarray}
R_{EH} = \sqrt{H(r_+)} = \sqrt{ \left( \frac{2 G m}{c^2} \right)^2 + \left( \frac{a_o c^2}{2 G m} \right)^2 }\,.
\label{loopEH}
\end{eqnarray}
Apart from $P$ terms, this is equivalent to Eq.~(\ref{quadEH}), asymptoting to the Schwarzschild radius for $m \gg M_P$ and the Compton wavelength 
for $m \ll M_P$ if we put $\beta = \sqrt{3} \gamma \zeta /4$. 

The important point is that central singularity of the Schwarzschild solution is replaced with another asymptotic region, so 
the black hole becomes a wormhole.
Metric~(\ref{g}) has three other important cosequences: (1) it permits the existence of black holes with $m \ll M_P$; (2) it implies a  duality between the $m < M_P$ and $m > M_P$ solutions; (3) it involves a unified expression for the Compton and Schwarzschild scales, with expression (\ref{loopEH}) suggesting the quadratic GUP. Further details can be found in Refs.~\cite{cmp} and \cite{poly}.

\section{GUP and Black Hole Thermodynamics}

Let us first recall the link between black hole radiation and the HUP~\cite{hawking}.  This arises because we can obtain the black hole temperature for $M \gg M_P$ by 
identifying $\Delta x$ with the Schwarzschild radius and $\Delta p$ with a multiple of the black hole temperature:
\begin{eqnarray}
kT = \eta c \Delta p = \frac{ \eta \hbar  c}{\Delta x} = \frac{\eta \hbar c^3}{ 2 G M} \, .
\label{temper}
\end{eqnarray}
This gives the precise Hawking temperature 
if we take $\eta = 1/(4\pi)$. 
The second equality in Eq.~(\ref{temper}) relates to the emitted particle and assumes that $\Delta x$ and $\Delta p$ satisfy the HUP. The third equality relates to the black hole and assumes that $\Delta x$ is the Schwarzschild radius. Both these assumptions require $M \gg M_P$ but the GUP and GEH suggest how they should be modified for $M \ll M_P$.

\begin{figure}[b]
\sidecaption
\includegraphics[scale=.35]{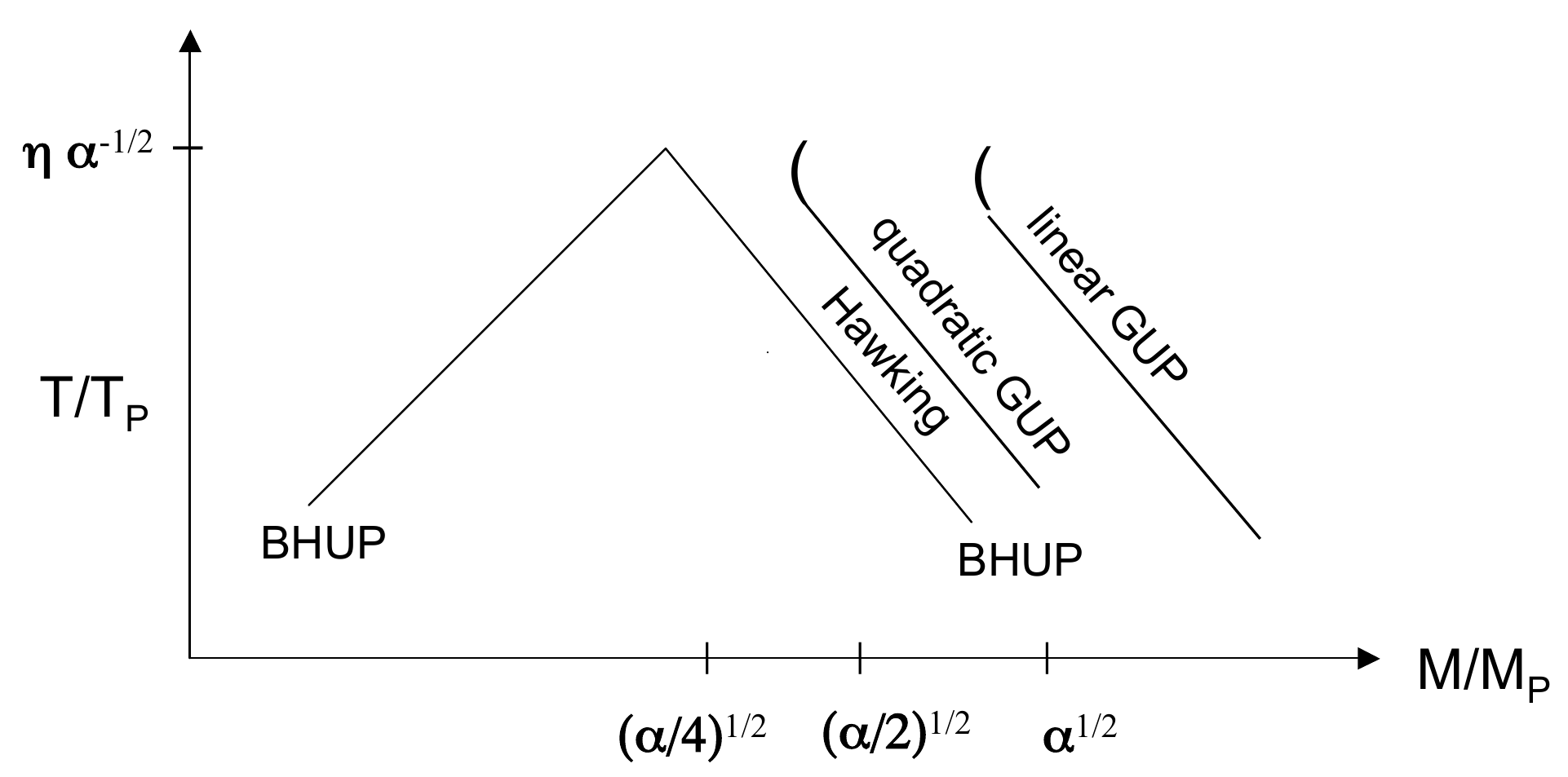}
\caption{Comparing black hole temperature predicted by Hawking, linear and quadratic GUP, BHUP correspondence.}
\label{temp}     
\end{figure}

Adler {\it et al.} \cite{Adler} calculate the modification required
if  $\Delta x$ is still 
associated  with the Schwarzschild radius but $\Delta p$ and $\Delta x$ are related by the linear form of the GUP. In this case, one obtains
a temperature 
\begin{equation}
T = {\eta M c^2 \over \alpha k} \left(1- \sqrt{1- \frac{\alpha M_P^2}{M^2}} \right) 
\approx
{\eta \hbar c^3 \over 2 G k M}  \left[ 1 + {\alpha M_P^2 \over 4M^2} \right] \, ,
\label{adlertemp2}
\end{equation}
where the last expression applies for $M \gg M_P$ and 
just represents a small perturbation to the standard Hawking temperature. However, as indicated in Fig.~\ref{temp}, the exact expression 
becomes complex when $M$ falls below $\sqrt{\alpha} \, M_P$, indicating a minimum mass. If we adopt the quadratic form of 
the relationship between $\Delta p$ and $\Delta x$, 
 the temperature becomes
\begin{equation}
T = {\sqrt{2} \, \eta M c^2 \over  \alpha k} \left(1 - \sqrt{1- { \alpha^2  \over 4} \left( {M_P \over M} \right)^{4} } \right)^{1/2} 
 \approx {\eta \hbar c^3 \over 2 G k M} \left[ 1 + \frac{ \alpha^2}{32} \left( {M_P \over M} \right)^{4} \right] \, ,
\label{looptemp2}
\end{equation}
so the deviation from the Hawking prediction is smaller than implied by Eq.~(\ref{adlertemp2}) but the exact expression still goes complex for $M < \sqrt{\alpha/2} \, M_P$. 
In either case, evaporation ceases at about the Planck mass.
So the GUP stabilizes the ground state of a black hole just as the HUP stabilizes the ground state of a hydrogen atom.

The BHUP correspondence suggests that the Alder {\it et al.}  argument must be modified since 
$\Delta x$  is given by Eq.~(\ref{loopEH}) rather than $2GM/c^2 $.
This 
makes a major qualitative difference for  $M \ll M_P$ because $\Delta x$ then scales as $M^{-1}$ rather than $M$ and this means that the temperature no longer goes complex.
As shown in Fig.~\ref{temp}, one obtains the {\it exact} solution \cite{cmp} 
\begin{eqnarray}
kT = \mathrm{min} \left[ \frac{\hbar \eta c^3}{2GM} \,  , \, \frac{2\eta Mc^2}{\alpha} \right]  \, .
\label{mess3}
\end{eqnarray}
The first expression is the exact Hawking temperature, 
with no small correction term.
However, one must cross over to the second expression below $M = \sqrt{\alpha /4} \, M_P$ in order to avoid the temperature going above the Planck value $T_P$.
The second expression
can be obtained by putting 
$\Delta x \approx \hbar/ (Mc)$ in Eq.~(\ref{temper}).
Since $T < T_P$ everywhere, the second equality still applies to a good approximation.
The different $M$-dependences for $M < M_P$ and  $M > M_P$ arise because
there are two different asymptotic spaces in  the LBH solution, 
corresponding to the $R$ and $r$ coordinates, 
so the quantity $\Delta x$ needs to be specified more precisely. 
Putting $r=2GM/c^2$
 implies 
$(\Delta x)_{R}/(\Delta x)_{r}   \approx
1$ for $M \gg M_P$ but
$(M/M_P)^{-2}$ for $M \ll M_P$.

Note that one can use another argument which gives a different temperature in the sub-Planckian regime. 
If the temperature is determined by the black hole's surface gravity \cite{hawking}, Eq.~(\ref{loopEH}) suggests 
$T \propto GM/R_S'^2 \propto 
M^3$ 
rather than $M$ for $M \ll M_P$.
The discrepancy arises because the temperature differs in the two asymptotic spaces by a factor $(M/M_P)^2$. The GUP argument only gives the temperature in the same space as the black hole event horizon, which is our space for $M>M_P$ but the other space for $M>M_P$. So the temperature of a sub-Planckian hole scales as $M^3$ in our space, as predicted by the surface gravity argument, and as $M$ in the other space, as predicted by the GUP argument \cite{cmp}.
Although there is no value of $M$ for which $T$ becomes zero,
there are still {\it effectively} stable relics since 
the temperature falls below the background radiation density  -- suppressing evaporation altogether -- below some critical mass and such relics might provide the dark matter~\cite{poly}.  

\section{Changing the Dimensionality}

\begin{figure}[b]
\sidecaption
\includegraphics[scale=.3]{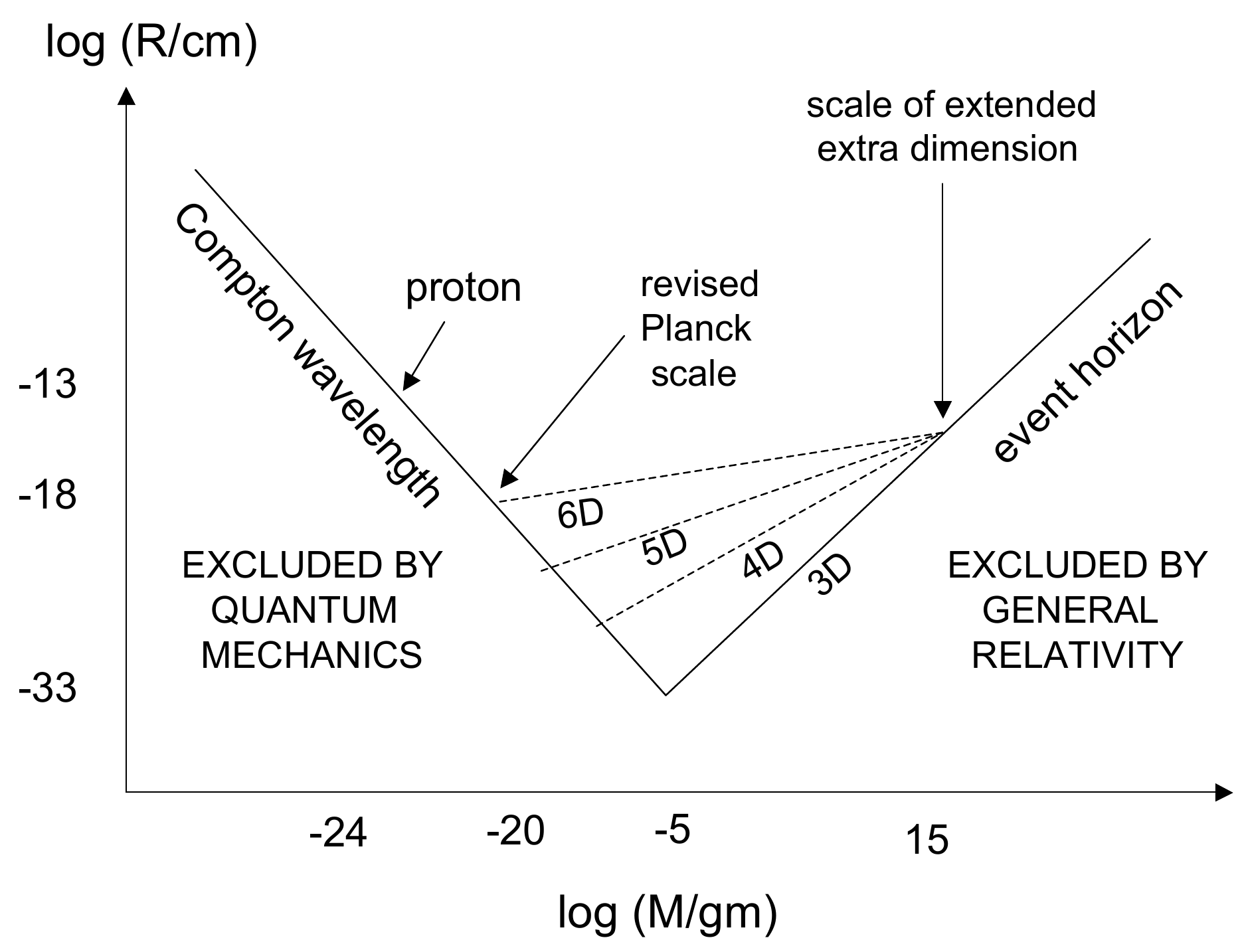}
\caption{Modification to Fig.~\ref{MR} for various numbers of spatial dimensions below some scale.}
\label{MR2}     
\end{figure}

The black hole boundary in Fig.~\ref{MR} assumes there are three spatial dimensions but
many theories suggest that the dimensionality could increase on small scales. Either the extra dimensions are compactified or matter is confined to a brane of finite thickness in the extra dimensions due to warping. In both cases, the extra dimensions are associated with some scale $R_C$. 
If there are $n$ extra dimensions and the black holes with mass below
$M_C = c^2 R_C/(2G)$ are assumed to be spherically symmetric in the higher dimensional space, 
then the Schwarzschild radius must be replaced with 
\begin{equation}
R_S =  R_C \left( \frac{M}{M_C} \right) ^{1/(n+1)} 
\label{higherBH}
 \end{equation}
for $M < M_C$, so the slope of the black hole boundary in Fig.~\ref{MR} becomes shallower, as indicated in Fig.~\ref{MR2} for various values of $n$. The new intersect with the Compton boundary just corresponds to the revised Planck scales.
We note that $R_S \propto M^{-1}$ for 2-dimensional holes ($n=-2$). This suggests some link with the idea that physics becomes 2-dimensional (rather than higher dimensional) close to the Planck scale \cite{mureika}, which offers an intriguing alternative interpretion of the BHUP correspondence.  

\end{document}